# Measuring Price Discovery between Nearby and Deferred Contracts in Storable and Non-Storable Commodity Futures Markets

Working Paper

Zhepeng Hu, Mindy Mallory, Teresa Serra, and Philip Garcia

Department of agricultural and Consumer Economics

University of Illinois, Urbana-Champaign

11/8/2017

# Measuring Price Discovery between Nearby and Deferred Contracts in Storable and Non-Storable Commodity Futures Markets

**Abstract**

Futures market contracts with varying maturities are traded concurrently and the speed at which they process information is of value in understanding the pricing discovery process. Using price discovery measures, including Putniņš' (2013) information leadership share and intraday data, we quantify the proportional contribution of price discovery between nearby and deferred contracts in the corn and live cattle futures markets. Price discovery is more systematic in the corn than in the live cattle market. On average, nearby contracts lead all deferred contracts in price discovery in the corn market, but have a relatively less dominant role in the live cattle market. In both markets, the nearby contract loses dominance when its relative volume share dips below 50%, which occurs about 2-3 weeks before expiration in corn and 5-6 weeks before expiration in live cattle. Regression results indicate that the share of price discovery is most closely linked to trading volume but is also affected, to far less degree, by time to expiration, backwardation, USDA announcements and market crashes. The effects of these other factors vary between the markets which likely reflect the difference in storability as well as other market-related characteristics.

**Measuring Price Discovery between Nearby and Deferred Contracts in Storable and Non-Storable Commodity Futures Markets**

Price discovery is a main function of commodity futures markets. Classic research on price discovery in agricultural futures markets focuses in three areas: determining which dominates, cash or futures price (Weaver and Banerjee 1982; Garbade and Silber 1983; Schroeder and Goodwin 1991); which of several geographically differentiated cash or futures markets dominates (Koontz, Garcia and Hudson 1990; Booth and Ciner 2001; Han, Liang, and Tang 2013; Janzen and Adjemian 2017); and whether there is a difference in the quality of price discovery in storable versus non-storable commodities (Gray and Rutledge 1971; Leuthold, Junkus, and Cordier 1989; Yang, Bessler, and Leatham 2001). The overwhelming evidence suggests that futures markets are the nexus of price discovery, and the cash markets play a lessor role as long as there is a liquid futures market available. Despite this evidence, we know little about where along the futures forward curve new information gets impounded into prices.

In futures markets, contracts are traded concurrently with several maturities per year. Knowing how each contract maturity contributes to price discovery is essential for market participants making sound hedging and trading decisions. The theory of price relationships along the forward curve for storable commodities is well developed by Working (1948, 1949). As explained in Tomek (1997), Working's theory for storable commodities is that deferred futures prices only make random adjustments to nearby futures prices and hold no additional value for price discovery, except in periods of backwardation when low inventories break down normal price linkages. For non-storable commodities, the economic theory of intertemporal price relationships is less developed.

Because non-storable commodities do not have arbitrage-enforced linkage through storage, different contract maturities are not as strongly linked as they are in futures for storable commodities (Gray and Rutledge 1971). Leuthold, Junkus, and Cordier (1989) argue that for non-storable commodities, futures contracts for different delivery months should have unique information value since they reflect market expectations of equilibrium conditions at different horizons.

Empirical studies find the nearby contract, on average, provide most price discovery in futures markets for agricultural commodities (Sanders, Garcia, and Manfredo 2008; Schnake, Karali, and Dorfman 2012), as well as other financial assets (Mizrach and Neely 2008; Chen and Tsai 2017). However, the reality of price discovery along the forward curve is likely to be more complex. The forward curve is constantly shifting as days to expiration decrease for each contract on the board and new contracts are added. As this happens, the nearby contract loses importance as the delivery period approaches, evidenced by falling volume and open interest. It is not known, however, how rapidly price discovery fades and whether it happens at the same rate as volume leaving the nearby contract. Knowing when the leadership of price discovery switches from the nearby contract to the next nearby contract is particularly important for contract rolling decisions that are made by both academic researchers and market participants.

The location of price discovery along the forward curve may also vary for a number of other reasons – some predictable and some unpredictable. USDA reports are released on a predetermined schedule and have been shown to impart important information about fundamentals that are quickly reflected in prices (Adjemian 2012; Lehecka 2014;

Lehecka, Wang, and Garcia 2014; Dorfman and Karali 2015; Mattos and Silveira 2016). While most studies only examine the nearby futures prices, it is possible that USDA reports also affect deferred futures prices as market participants may use information in the release to make medium- and long-term trading and hedging decisions. Moreover, price discovery is particularly important in periods of price turbulence when market participants face higher risks. When periods of extreme price volatility arise, price discovery may shift as liquidity conditions and market participants' expectations change. Recent booms and busts in commodity prices have raised concerns on the price discovery function of commodity futures markets. While most studies have focused on whether price discovery of agricultural commodity futures markets has been harmed by financial speculations (Irwin 2013; Janzen et al. 2014; Bruno, Büyükşahin, and Robe 2016), the effect of price turmoil on price discovery along the forward curve has not been examined.

This paper measures price discovery between nearby and deferred futures each day from 2008 to 2015 using transactions data for corn and live cattle that are time-stamped to the second from CME Group's Top-of-Book dataset. The two commodities provide a useful comparison because Corn and Live Cattle contracts are storable and non-storable, respectively. The period examined is characterized by advances in electronic trading that have dramatically changed the landscape of agricultural commodity futures markets (Irwin and Sanders 2012). Some contracts, like corn, have flourished and others, like live cattle, have floundered in terms of public trust and measures of market quality like volume and volatility (Gee 2016). Despite the importance of the transition to electronic trading, however, little work has been done to update the extant literature on price discovery in agricultural commodity markets. The period examined also includes periods when corn and live cattle prices were extremely volatile, as well as when markets were in backwardation, which allows us to identify how price discovery along the forward curve changes in different market situations.

This article is the first to use high-frequency data for studying price discovery between nearby and deferred contracts in electronic agricultural commodity futures markets. Previous studies typically used data at a daily frequency, which did not permit a dynamic characterization of how futures price discovery rolls from one contract to the next as the nearby nears expiration. We employ a recently developed price discovery metric, information leadership share (Putniņš 2013), which is designed to be used on high frequency data samples. While previous research has focused on price forecasts of nearby and deferred futures (Sanders, Garcia, and Manfredo 2008; Schnake, Karali, and Dorfman 2012), the information leadership share enables us to directly measure the relative proportion of information impounded in nearby and deferred futures prices. In addition, compared to conventional price discovery measures, this measure is more robust to differences in noise levels that appear in nearby and deferred futures prices.

We begin our analysis by documenting patterns in daily price discovery shares between nearby and deferred futures in the whole sample period as well as in the nearby period of each contract month. Subsequently, we conduct regression models to investigate determinants of price discovery between nearby and deferred futures. Findings suggest the nearby contract dominates deferred contracts in price discovery when it has more trading volume, which typically happens before entering the delivery period. The nearby contract plays a more important role in terms of price discovery in the corn market than in the live cattle market. Differences in behavior between the corn and live cattle markets also occur in response to USDA announcements, and in periods of backwardation and crashes.

**Price discovery measures**

Garbade and Silber (1983) were the first to develop a measure (the $GS$ measure) to quantify price discovery between closely linked prices that share the same fundamental value. Garbade and Silber propose the following model of price behavior:

(1) $$\begin{bmatrix} p_{1,t} \\ p_{2,t} \end{bmatrix} = \begin{bmatrix} \alpha_1 \\ \alpha_2 \end{bmatrix} + \begin{bmatrix} 1 - \beta_1 & \beta_1 \\ \beta_2 & 1 - \beta_2 \end{bmatrix} \begin{bmatrix} p_{1,t-1} \\ p_{2,t-1} \end{bmatrix} + \begin{bmatrix} \omega_{1,t} \\ \omega_{2,t} \end{bmatrix}$$

where $p_{1,t}$ and $p_{2,t}$ are the prices for nearby and deferred futures contracts at time $t$, repectively. The coefficients $\beta_1$ and $\beta_2$ are the effect of one period lagged price for the deferred contract on the current price for the nearby contract and vice versa, respectively. The shares:

(2) $$GS_1 = \frac{\beta_2}{\beta_1 + \beta_2}, \quad GS_2 = \frac{\beta_1}{\beta_1 + \beta_2}$$

are used for measuring the proportional contribution of each contract to the price discovery process. If $\beta_1 = 0$, then $GS_2 = 0$ and $GS_1 = 1$, and the results can be interpreted as 100% of the new information is reflected first in the nearby contract. The coefficients in equation (1) can be estimated via the OLS by rearranging equation (1) algebraically as follow:

(3) $$\begin{bmatrix} \Delta p_{1,t} \\ \Delta p_{2,t} \end{bmatrix} = \begin{bmatrix} \alpha_1 \\ \alpha_2 \end{bmatrix} + \begin{bmatrix} \beta_1 \\ -\beta_2 \end{bmatrix} [p_{1,t} - p_{2,t}] + \begin{bmatrix} \omega_{1,t} \\ \omega_{2,t} \end{bmatrix}.$$

Applications of the *GS* measure, particularly in the area of commodity cash and futures markets, often suggest economically reasonable results that futures lead the cash price (Oellermann, Brorsen, and Farris 1989; Schroeder and Goodwin 1991). However, from an econometric viewpoint, the *GS* measure is derived from lead-lag regressions which ignores the possibility that the two prices share a common stochastic trend that represents the common efficient price. The advances in multivariate cointegration and error correction modeling introduced by Engle and Granger (1987) made it possible to constrain multiple prices to share a common efficient price in a cointegration framework. Hasbrouck (1995) information share (*IS*) and Harris-McInish-Wood (2002) component share (*CS*)[1] are the most widely used price discovery measures based on cointegration and error correction model. Without losing the functional generality of the lead-lag regression approach of the *GS* measure, both *IS* and *CS* depend on the notion that prices for different contracts (or in different markets) for the same commodity (or other asset) can deviate from each other in the short term, but will converge to their common fundamental value in the long term.

Assume the fundamental value of a commodity follows a random walk:

(4) $$w_t = w_{t-1} + \mu_t, \quad \mu_t \sim N(0, \sigma_\mu)$$

where $w_t$ is the fundamental value at time $t$, and $\mu_t$ is *i.i.d*. The observed futures price $p_{i,t}$ at time $t$ tracks the fundamental value $\delta_i$ periods lagged can be expressed as:

(5) $$p_{i,t} = m_{t-\delta_i} + s_{i,t}, \quad s_{i,t} \sim N(0, \sigma_{s_i})$$

where $i = 1$ and 2 stands for nearby and deferred contracts, and $s_{i,t}$ is *i.i.d* and uncorrelated between nearby and deferred contracts. Thus, price deviations from the fundamental value are only transient and the prices for nearby and deferred contracts are cointegrated.

Both *IS* and *CS* are derived by estimating the following (bivariate) VECM:

(6) $$\Delta \boldsymbol{p}_t = \boldsymbol{\alpha}(\boldsymbol{\beta}'\boldsymbol{P}_t - \mu) + \sum_{j=1}^{J} \boldsymbol{\Gamma}_j \Delta \boldsymbol{p}_{t-1} + \boldsymbol{e}_t$$

where $\boldsymbol{p}_t = (p_{1,t}, p_{2,t})'$ is a vector of futures prices for nearby and deferred futures contracts at time $t$, respectively. The $\boldsymbol{\beta} \in \mathbb{R}^2$ is a cointegrating vector. Following Figuerola-Ferretti and Gozalo (2010), we relax the assumption of $\boldsymbol{\beta} = (1, -1)$ that is formally used in the literature and specify a constant term $\mu$ in the long-run equilibrium to incorporate carrying charge. The parameter vector $\boldsymbol{\alpha} = (\alpha_1, \alpha_2)'$ contains error correction coefficients that measure the adjustment speed at which violations of the long-run price equilibrium are corrected. The vector $\boldsymbol{\Gamma}_j \in \mathbb{R}^{2\times 2}$ is a vector of coefficients for the autoregressive terms representing short-run dynamics and $J$ is the number of lags included in the model. The error term $\boldsymbol{e}_t$ is a zero-mean vector of white noise residuals with the covariance matrix:

(7) $$\Sigma = \begin{pmatrix} \sigma_1^2 & \rho\sigma_1\sigma_2 \\ \rho\sigma_1\sigma_2 & \sigma_2^2 \end{pmatrix}.$$

Harris, McInish, and Wood (2002) show that the *CS* can be calculated from the normalized orthogonal to the vector of error correction coefficients, $\alpha_\perp = (\gamma_1, \gamma_2)'$. By noting that $CS_1 + CS_2 = 1$,

(8) $$CS_1 = \gamma_1 = \frac{\alpha_2}{\alpha_2 - \alpha_1}, CS_2 = \gamma_2 = \frac{\alpha_1}{\alpha_1 - \alpha_2}$$

are the *CS* measures for nearby and deferred contracts, respectively.

In Hasbrouck (1995) and other studies, the *IS* measures can be estimated using the Vector Moving Average (VMA) representation of the VECM. However, Baillie et al. (2002) show that the *IS* can be directly derived from the error correction coefficients and the variance-covariance matrix of the error terms, which makes the estimation much easier. Given the Cholesky factorization of the VECM residual covariance matrix, $\Sigma = MM'$, where

(9) $$M = \begin{pmatrix} \sigma_1 & 0 \\ \rho\sigma_2 & \sigma_2(1-\rho^{1/2})^{1/2} \end{pmatrix} = \begin{pmatrix} m_{11} & 0 \\ m_{12} & m_{22} \end{pmatrix}.$$

The *IS* measures for nearby ($IS_1$) and deferred ($IS_2$) contracts can be calculated using:

(10) $$IS_1 = \frac{(\gamma_1 m_{11} + \gamma_2 m_{12})^2}{(\gamma_1 m_{11} + \gamma_2 m_{12})^2 + (\gamma_2 m_{22})^2}, IS_2 = \frac{(\gamma_2 m_{22})^2}{(\gamma_1 m_{11} + \gamma_2 m_{12})^2 + (\gamma_2 m_{22})^2}.$$

Essentially, the *IS* measures each price series' relative contribution to the variance of the innovations to the common factor. Following Baillie et al. (2002) and many others, we calculate the *IS* measures using each of the two alternative orderings of price series and then take the average to avoid the effect of the ordering of the variables in the VECM.

In the conventional sense, price discovery metrics measure the leadership in impounding new information. The majority of price discovery studies consider that a price series dominates price discovery if it incorporates new information about the fundamental value faster. Recent studies (Yan and Zivot 2010; Putniņš 2013) have showed that *IS* and *CS* are only consistent with the "who moves first" view of price discovery when the levels of noise in the price series are similar. In particular, using *IS* and *CS* may lead to overstating the price discovery contribution of the less noisy contract.

Yan and Zivot (2010) employ a structural cointegration model to analytically demonstrate what exactly *IS* and *CS* measure. They show that the *IS* measures a combination of leadership in impounding new information and the relative level of avoidance of noise, while the *CS* merely measures the relative response to transitory frictions. Yan and Zivot (2010) propose to use a combination of *IS* and *CS* such that the response to transitory frictions can be net out. The measure of the relative impact of a permanent stock, developed by Yan and Zivot (2010) and termed "information leadership" (*IL*) in Putniņš (2013), is expressed as follows:

$$(11) \quad IL_1 = \left|\frac{IS_1}{IS_2}\frac{CS_2}{CS_1}\right|, IL_2 = \left|\frac{IS_2}{IS_1}\frac{CS_1}{CS_2}\right|$$

where $IL_1$ and $IL_2$ are the *IL* measures for nearby and deferred contracts, respectively. The *IL* is not a "share"; to make it comparable to *CS* and *IS* and easy to interpret, Putniņš (2013) defines the information leadership shares for nearby ($ILS_1$) and deferred ($ILS_2$) as:

$$(12) \quad ILS_1 = \frac{IL_1}{IL_1 + IL_2}, ILS_2 = \frac{IL_2}{IL_1 + IL_2}.$$

Using a simulation study, Putniņš (2013) shows that *IS* and *CS* are biased when price series differ in noise levels because they both jointly measure relative speed in reflecting innovations in the fundamental value and relative avoidance of noise, while the *ILS* is robust to differences in noise levels. The levels of noise in prices for nearby and deferred futures contracts can be very different due to differences in volume and bid-ask spread (Wang, Garcia, and Irwin 2013). Therefore, the *ILS* is more appropriate than *IS* and *CS* for cointegrated nearby and deferred futures prices. Thus, when prices are cointegreated series, we use the *ILS* measure.

In this paper, we examine the relationships between nearby and deferred futures contracts in a bivariate context. This is consistent with the literature which focuses on the relationship between two markets at the same time and then assesses these relationships. It is also consistent with the *ILS* measure which is based on the assumption of only two price series (Yan and Zivot, 2010; Putniņš 2013). Finally, the use of bivariate analysis reduces the likelihood that the price discovery process is masked by a high degree of price interaction and possible multiple stochastic trends which could emerge across similar contract prices.

**Data**

The analysis centers on the corn and live cattle futures contracts traded at the Chicago Mercantile Exchange (CME). These markets represent the most actively traded storable and non-storable agricultural commodities, respectively. The sample period studied for corn is from January 14, 2008 through December 14, 2015, and the period used for live cattle ranges from January 1, 2008 to December 31, 2015. The period examined is characterized by the growing relevance of electronic trading in agricultural commodity futures markets. The electronic platform's shares of corn and live cattle futures trades were about 80% and 10% at the beginning of 2008 (Irwin and Sanders 2012), and both rose to over 95% in 2015 (Gousgounis and Onur 2016). The period examined also includes boom-bust cycles in corn and live cattle prices, as well as periods when the markets were in backwardation.

We use high frequency trade data obtained from the CME Group's Top-of-Book database. Our data include transaction prices that are time stamped to the second and ordered chronologically by sequence numbers. We aggregate the data to one second frequency by taking the first transaction when multiple transactions have the same time stamp. If there is no transaction within a second, we create an entry for that second with the most recent transaction information. The benefit of using a short sampling interval is that it helps eliminate the contemporaneous correlation between innovations in nearby and deferred prices, so that the sequence of price response can be better identified (Hasbrouck 1995). The CME electronic trading system (Globex) is open nearly 24 hours, however, we only consider the day-time trading session of both the corn and live cattle contracts when the most active trading occurs.

Corn futures market has five delivery months: March, May, July, September, and December. Live cattle futures have six delivery months: February, April, June, August, October, and December. For both corn and live cattle futures, multiples contracts with different maturity dates are traded each day. Because volumes in the far-dated deferred contracts are quite low, we use the first five nearby contracts for corn, and refer to them as the nearby, deferred 1, deferred 2, and so on. For live cattle futures, the first four nearby contracts are used, and we refer to them similarly in the rest of this paper. Corn futures contracts expire on the business day prior to the 15th calendar day of the delivery month and live cattle futures contracts expire on the last business day of each maturity month. We define a contract to be the nearby from the business day after the previous nearby contract expires through the current nearby contract expiration. We refrain from rolling the nearby contract to the next to the last possible date so we can more clearly identify how price discovery share in the nearby declines as expiration approaches.

**Empirical results**

Since the *ILS*, as well as *CS* and *IS*, are based on a VECM, we test for cointegration first. Cointegration often characterizes nearby and deferred futures price relationship when sampling at a daily frequency. However, intraday prices for nearby and deferred futures may not be cointegrated. For example, intraday nearby and deferred futures prices for a storable commodity may not be cointegrated during days when the market is in backwardation and prices for different maturities are not linked by storage. For a non-storable commodity, prices for different maturities are not linked by storage and arbitrage and therefore may not be cointegrated within the day.[2]

We employ Johansen tests to assess cointegration between the nearby and each deferred contract on a daily basis. Because prices can vary very little when the market is tranquil or if there is a price limit move, we exclude days in which price updates are too infrequent to enable us to perform statistical tests or estimation. The days excluded from the analysis account for about 1.8% of the sample for corn. For live cattle, we have about 3% of the total trading days excluded for the first and second contract pairs, and about 10% for the third contract pair. For corn futures, nearly all of the excluded days are in the period of 2008 through 2010 when the share of electronic trading was relatively small. We have more days excluded for live cattle because it is a much less active market than the corn market.

Based on the Johansen cointegration rank test, we group our data into three categories[3]: 1) Stationarity: intraday nearby and deferred futures prices are both stationary $I(0)$ series, in which case we fail to reject the null of hypothesis of a rank of 2 at the 5% significance level. 2) Cointegration: intraday nearby and deferred futures prices are cointegrated $I(1)$ series, in which case we fail to reject the null hypothesis of a rank of 1 at the 5% significance level. 3) Non-cointegration: intraday nearby and deferred futures prices are both not stationary and they are not cointegrated, in which case we fail to reject the null hypothesis of a rank of 0 at the 5% significance level.

Table 1 summarizes the percentage of days that belong to each category for each contract pair. In both markets, the probability of both prices being stationary is slightly above 20%, with the exception of 27% for the nearby and deferred 1 contract pair for corn. The percentage of the days in which we reject the null hypothesis of no cointegration is about 70% across all contract pairs for both commodities, suggesting nearby and deferred futures share the same stochastic trend most of the time. The percentage of non-cointegration days ranges from 3% to 5% for corn and from 7% to 13% for live cattle, which likely reflects the difference in storability. Results also suggest that the percentage of non-cointegration generally increases at more deferred contracts. This is somewhat expected since contracts at more distant maturities may reflect different price information.

Figure 1 and figure 2 present the distribution of Johansen test results through time for corn and live cattle, respectively. Each observation is colored coded to reflect the cointegration test results, and located relative to the vertical axis to represent the volume share of the nearby contract. Volume share is the volume of the nearby contract divided by the total volume of the nearby and deferred contracts on the same day. Shaded areas represent backwardation periods. In both figures, we see volume share presents a cyclic pattern, with the nearby contract's volume share decreasing as expiration approaches and then increasing with movement to the next nearby contract. In figure 1, a clear pattern emerges for corn, with nearly all of the non-cointegration days (red squares) appearing in backwardation periods. This result is consistent with our expectation that periods of backwardation will cause the link between maturities to deteriorate in storable markets. In contrast, figure 2 shows non-cointegration in the live cattle market does not necessarily concentrate in backwardation periods, which is consistent with live cattle's non-storable character. In both markets, we find that as expiration approaches the number of non-cointegration days increases. Further, nearby and deferred futures prices are less likely to be both stationary in the first few weeks after entering the nearby period in both markets[4].

*Price discovery shares for the nearby contract relative to deferred contracts*

Since the *ILS* is based on cointegration, we calculate *ILS* (equation (12)) for each day when intraday transaction prices for nearby and deferred contracts are $I(1)$ and cointegrated. For days when intraday prices for nearby and deferred futures are both stationary, we use the *GS* measure (equation (2)). Following a bivariate context, we calculate daily price discovery shares for the nearby contract relative to each deferred contract separately. For the *ILS*, we select the number of lags of the VECM based on the Bayesian Information Criteria (BIC), and lags between 1 and 10 lags were common for both commodities. Following Garbade and Silber (1983), we estimate equation (3), and when calculating the GS measure we set negative estimates of $\beta_1$ and $\beta_2$ to 0 since negative values have no conceptual meaning.

To begin we plot daily price discovery shares for the nearby contract and each deferred contract. Figures 3 and 4 present the results for corn and live cattle, respectively. In figure 3 and to a lesser extent in figure 4, the nearby contract appears to dominate price discovery except when it is near expiration. The gradual decline in the price discovery share, which is more apparent in the corn market, is similar to the behavior of volume share presented in figures 1 and 2. Examination of the relationships for each commodity also suggests that a dominance of the nearby contract relative to distant contracts particularly in the corn market. The nearby price discovery share is progressively more concentrated near the top of the *ILS* range at more deferred contracts. Table 2, which reports the averages of daily price discovery (*ILS* and *GS*) and volume shares for the nearby contract and deferred contracts for corn and live cattle, provides a clearer picture of this behavior. In both markets, price discovery and volume shares for the nearby contract increase as the time to the deferred contract increases. This term structure is expected because volume and accompanying liquidity at distant horizons are usually lower which implies less market information.

For corn futures, the nearby contract only slightly dominates the first deferred contract in price discovery with an average price discovery share of 52%. However, the nearby price discovery share rises quickly as the time to the deferred contracts increases. By the deferred 4 contract, the nearby price discovery share has reached 78%. In general, these findings are consistent with Working's theory for storable commodities that less price discovery exists outside the nearest maturity.

Compared to corn futures, the nearby live cattle contract is less dominante in price discovery. On average, the nearby contract does not provide more price discovery than the next nearby contract with an average price discovery of 42%. Compared to the second and the third deferred contracts, the nearby contract contributes by about 56% and 64% of the price discovery, respectively. Informatively, volume shares for deferred 1 contract are appreciably below 50%, and only reach 50% for the deferred 2 contract, suggesting much less liquidity in the nearby contract which again contrasts rather sharply with the corn contract. For live cattle futures, since there is no storage arbitrage to link the contracts with differing maturities, contracts for different delivery dates provide information of equilibrium conditions at different future dates (Leuthold, Junkus, and Cordier 1989). Thus, the difference in the dominance of the nearby contract for corn and live cattle attribute to their difference in storability.

*Price discovery shares in the nearby period for each contract month*

To examine more closely the behavior of price discovery, we plot for each contract average daily nearby price discovery and volume shares relative to the next nearby contract (figures 5 and 6). The horizontal-axis in all plots is the trading days to contract expiration. Because the December corn contract becomes important for hedging and pricing early in the marketing year, we also compare the nearby July contract to the December contract (panel 6, figure 5).

Price discovery shares for corn are presented in figure 5 and they exhibit similar patterns for most contract months, except when September is the nearby contract. Initially, most nearby contracts have a price discovery share around 80% and continue to dominate (i.e. share higher than 50%) price discovery until 2-3 weeks prior to contract expiration. In general, the price discovery share moves in tandem with the volume share, declining sharply as trading volume decreases in the nearby contract. The nearby contract loses its dominant role in price discovery nearly at the same time as it loses dominance in volume share. This typically happens 2-3 weeks prior to contract expiration which roughly coincides with the beginning of the delivery window.

The notable exception to the price discovery pattern described is the September contract (figure 5, panel 4). While most contracts begin with a price discovery share at nearly 80%, the September contract, which is compared to the December contract, only initially and briefly breaks 50%, and then remains well below 40% to expiration. In effect, this says that the September contract plays little if any role in price discovery. To investigate the importance of the December contract, we examined the price discovery and volume shares for the July contract relative to December contract (figure 5, panel 6). This exhibits a pattern more typical of the price discovery patterns in the other corn panels. These findings suggest that as the July contract approaches expiration, price discovery shifts rapidly to the December contract. The relative lack of importance of the September contract in the pricing process is likely due to its position between two crop years where its prices reflect mixed information relative to the new crop. Other researchers have observed related differences in the September contract in its volatility patterns (Smith 2005) and ability to predict subsequent cash prices (Leath and Garcia 1983). Since the December contract reflects information on the new crop which becomes highly relevant in the summer months, and its price is the focus of many hedgers and market participants, its leadership in price discovery for such an extended period is understandable.

Figure 6 plots the shares for the live cattle market. Both price discovery and volume shares exhibit a similar general downward trend across all contract months, albeit the shares follow each other less closely as expiration approaches. The volume share decreases almost linearly through the period, and the nearby contract loses its pricing dominance almost at the same point when volume share declines below 50%. Both the dominance of price discovery and trading volume of contracts switch to the next nearby contract about 5-6 weeks prior to expiration, which is much earlier than in the corn market. For some contract months like June, August and October, the price discovery share increases in the final trading days and even exceed 50% in the last 1 week prior to expiration [5]. The volatility of price discovery shares in the last few days reflects the unstable intertemporal price relationships near expiration in the live cattle market that has been widely identified in the literature (e.g. Leuthold 1972; Naik and Leuthold 1988; Liu et al. 1994).

*Determinants of price discovery between nearby and deferred contracts*

In this section, we assess the ability of several factors to explain the nearby contracts' price discovery share (*PS*) relative to deferred contracts using a regression framework. The analysis focuses on equations (13) and (14) which identify the structure of the relationships for the corn and live cattle markets:

(13) $PS_{corn,d} = b_0 + b_1 Volumeshare_d + b_2 Expiration_d + b_3 Expiration^2_d +$

$b_4 Backwardation_d + b_5 WASDE\&CP_d + b_6 Grainstocks_d +$

$b_7 Crash_d + b_8 Stationarity_d + \varepsilon_d,$

and

(14) $PS_{livecattle,d} = b_0 + b_1 Volumeshare_d + b_2 Expiration_d + b_3 Expiration^2_d +$

$b_4 Backwardation_d + b_5 CF_{d-1} + b_6 Crash_d +$

$b_7 Stationarity_d + \varepsilon_d.$

*PS* is the daily price discovery share on a specific date *d* for the nearby contract relative to a deferred contract. *Volumeshare* and *Expiration* are the nearby contracts volume share, and a count variable that identifies the number of days to expiration. Similar factors were identified in the VIX (Chen and Tsai 2017) and bond futures markets (Mizrach and Neely 2008; Fricke and Menkhoff 2011). To allow for the non-linear behavior observed in the price discovery share near expiration, we include a quadratic term, $Expiration^2$.

To measure the impact of backwardation on the price discovery process, we create a dummy variable $Backwardation$ that equals one on days in which deferred futures settlement price was below the nearby futures settlement price. For corn, we have shown that the breakdown in cointegration between nearby and deferred prices is most likely to occur when the market is in backwardation. However, it is not clear which contract dominates price discovery when the market is in backwardation. For live cattle futures, we expect the presence of backwardation to have no impact since live cattle are non-storable and thus may not play a role in the pricing process. Because we use a combination of *GS* and *ILS* as our price discovery measure, it is possible that the magnitude of the price discovery share on a given day is affected by the price discovery measure applied. Thus, we introduce a dummy variable $stationary$ that equals one if the prices are both stationary in which case the *GS* is used as the measure of price discovery.

In addition to the factors identified in the earlier part of our analysis, we include two market-related factors which may influence the degree to which one contract responds more rapidly than another. The first is USDA market reports. Price discovery in agricultural commodity futures markets is affected by USDA market reports that carry important information on market fundamentals. Garcia and Leuthold (1992) examine the effect of USDA crop production reports on harvest and deferred contracts. While they find little difference in the estimated effects, daily settlement prices were used which may not reflect the intraday movements in price. For the corn market, we consider three important USDA grain market reports: the World Agricultural Supply and Demand Estimate report (WASDE), Crop Production report and Grain Stocks report. Usually, the WASDE and Crop Production reports are released in the second week of each month. The USDA started to release the WASDE and Corp production reports during regular trading hours after July 2012. Before this time, these two market reports were released before market opening. Because these two reports are released on the same day, we create a single dummy variable $WASDE\&CP$ for the two reports. The Grain Stocks report releases quarterly and release days are usually in mid-January, and the end of March, June and September. In our sample period, Grain Stocks reports are released at 12 p.m. during the regular trading hours. Similarly, we use a dummy variable $Grainstocks$ for Grain Stocks reports. For the live cattle market, we use the Cattle on Feed report that has the largest impact on the live cattle market among all USDA reports (Isengildina, Irwin, and Good 2006). During our sample period, Cattle on Feed reports are released on the third Friday of each month after day time trading session. We create a dummy variable $CF$ for Cattle on Feed reports. Grain market reports are released either before market opening or during the regular trading session, thus we would expect to observe any impacts on the release day. In contrast, Cattle on Feed reports are released after regular trading hours and their effects would most likely be observed on the next trading day. Therefore, the dummy variables for the grain market reports equal to 1 on the release date and the dummy variable for Cattle on Feed reports equals 1 on the following trading day.

A second factor is related to the events which occurred in our sample. Both markets experienced dramatic declines in prices which may have changed price discovery along the forward curve. To examine this effect, we introduce a dummy variable $Crash$ that equals to one in the market crash period. The corn market crash period is defined from July 03, 2008 when corn prices peaked, to December 08, 2008 when prices bottomed out in the 2008 great recession. For live cattle, the recent 2015 collapse in cattle prices has caused concerns about the price discovery function of the live cattle futures market. Hence, we use the entire year of 2015 as the crash period.

While the effects of USDA reports and the Crash periods may be uncertain, their inclusion should provide insights into how markets respond to large changes in information. Based on our earlier analysis, one would anticipate that the effect would be registered first in the market with highest volume traded and liquidity, but it may also be influenced by the timing of the release of the report and by users of the information.

Regression models are estimated separately for each pair of contracts in the corn and live cattle markets. Newey-West standard errors that correct for serial correlation and heteroscedasticity are reported with coefficient estimates. The lag length used for the Newey-West estimator is selected using the method described in Newey and West (1994). Adjusted *R*-squared values and number of observations for each equation are reported in the lower panel in each table.

Table 3 and table 4 report the regression results for the two markets. The adjusted *R*-squared values in the corn models are consistently higher than in the live cattle models, reinforcing the graphical analysis which demonstrated that price discovery in the corn market is more systematic than in the live cattle market. Within each market, *R*-squared values decrease as the length of time between the nearby and deferred contracts increases, indicating the model fits better for contract pairs with closer maturities.

In general, the coefficients of the principal factors have expected signs for both commodities. As anticipated, the coefficient of volume share is significant and positive across contract pairs in both markets, which is consistent with the economic intuition that higher trading volume facilitates information processing in terms of both speed and amount. Among all the variables, volume share has the largest influence in both markets. A 1% increase in the nearby contract's volume share increases the nearby contract's share of price discovery by about 0.6% to 0.7% relative to deferred contracts in the corn futures market, and about 0.3% to 0.4% in the live cattle market, after controlling for the other factors.

In the corn market, the coefficients of the days to expiration variable and its quadratic term are both significant and positive, which is consistent with our early observations in figures 3 and 5. As expiration approaches, price discovery first declines gradually and then drops off more sharply. The relationship between days to expiration and price discovery share is less pronounced in the live cattle market. The days to expiration variable in the nearby and deferred 1 contract pair is significant and negative and its quadratic term has a significantly positive coefficient. The estimated coefficient of $Expiration^2$ in the live cattle nearby and deferred 1 contract pair is 0.00033 (to the fifth decimal place), suggesting the overall effect of days to expiration is positive only when the number of days to expiration is roughly larger than 24[6]. This evidence supports our early observation in figure 6 that the price discovery share for the live cattle nearby contract tails up in the last few weeks in nearby periods. The overall days to expiration effect is not statistically significant in the other two contract pairs.

The presence of backwardation only has significant effect on price discovery between nearby and deferred 1 contract in the corn market. The corn nearby contract's share of price discovery relative to deferred 1 contract increases about 7% relative to the first deferred contract when the prices are in backwardation. Significant backwardation effects do not emerge in the other corn contract pairs. As expected, coefficients of the backwardation variable in the live cattle market are not significant, reflecting the non-storability of live cattle.

The coefficients for the stationarity dummy variables have mixed signs, pointing to slightly faster incorporation of information in the nearby corn contracts when using the *GS* measure, but slightly slower incorporation of information in the nearby cattle contract. The reason for this difference is not clear, but although several of the coefficients are significant, the size of the coefficients is small (less than 4%). Hence, using both price discovery measures together has little influence on the findings of the analysis.

The effect of USDA reports also differs in the two markets. In the live cattle market, Cattle on Feed reports have no significant influence on the price discovery across all contract pairs. In the corn market, some negative significant effects of the USDA grain market reports on contract pairs emerge in the more deferred contracts. The limited USDA effects observed in most of the contract pairs may arise for at least two reasons. One explanation is that the effect of an information release in electronic markets is very quickly incorporated into prices and then disappears (Lehecka, et al., 2014). This suggests that any differences in information processing can be masked by market activity in the rest of the day. Alternatively, the information is just quickly incorporated in the different contracts. The significant negative effects at more distant contracts may reflect the observation that many large information-changing releases are often USDA crop production reports which are more likely to influence deferred new crop corn contracts.

The effect of the crash factor is also mixed in the two markets. In the cattle market, the nearby contract's contribution to the price discovery decreased progressively from 4% to 7% as the time between contracts increased in 2015. Part of the decline may have resulted from a loss of confidence by some traders and market participants in nearby contracts which were not heavily traded to start (table 2). Additionally, the live cattle futures market heavily reflects the activities of feed lot operators who in the face of sharp price declines may have been quick to assure their positions in deferred contracts. In the corn market, the response to the July-early December crash in prices appears more direct. Nearby and the deferred 1 and 2 contracts responded in a similar manner to the sharp decline in prices. At more distant horizons, the nearby contracts which were most closely tied temporally to the initial sharp decline and possessed greater liquidity responded more quickly. Since the September contract which is normally not a leader in price discovery (figure 5) falls into the deferred 1 and 2 contract observations during this period suggests that the strength of the great recession likely moved all of the more nearby prices quickly and in a similar manner.

**Discussions and conclusion**

Understanding price discovery along the futures forward curve is important for market participants in making sound trading, hedging and production decisions. In the corn and live cattle futures markets, we quantify price discovery using intraday data, and graphical and statistical analysis for the 2008-2015 period—a period of highly volatile prices and a movement to electronic trading. The use of intraday data allows us to develop daily measures of price discovery and permits a more detailed assessment of the pricing process. We measure the price discovery share between nearby and deferred contracts, estimating the relative importance of nearby and deferred contracts in the pricing process, identifying when the dominance of price discovery switches from contract to contract, and estimating the relative importance of the factors that influence price discovery.

Our results demonstrate important differences and similarities between price discovery in the corn and live cattle markets. Many of the differences are related to storability which allows contract prices to be more closely linked through arbitrage. Price discovery is more systematic in the corn than the live cattle market, and resides more deeply, particularly in the corn market, in nearby as opposed to deferred contracts. The fact that nearby contract in the live cattle futures market plays a less dominant role in price discovery suggests that it may be misleading to rely solely on it for pricing signals. In both markets a clear term structure in price discovery appears. The share of price discovery is larger in the nearer to maturity contracts.

In both markets the relative amount of new information reflected in the nearby contract decreases as expiration approaches and trading becomes less active. In the corn futures market, the price discovery share for the nearby contract begins high and stable before declining and falling below 50% about 2-3 weeks before contract expiration. The exception is the September corn contract which rarely dominates the next nearby (December) contract. This finding suggests that market participants who require information on pricing should begin to follow the December contract as early as the beginning of July. In the live cattle market, nearby contracts lose their dominance to the next nearby contract as early as 5 to 6 weeks prior to their expiration.

Graphical analysis of price discovery measures reveals a close relationship between price discovery and trading volume. The nearby contract loses its dominance in price discovery to the next nearby contract when it is less actively traded, which identifies the importance of trading volume as it reflects market's opinion on pricing. This result leads to a practical recommendation for researchers and practitioners regarding how to roll contracts when creating a continuous series of nearby prices. Since price discovery dominance is closely linked to trading volume share, we recommend rolling to the next nearby contract when it achieves larger than 50% of the volume share.

Regression analysis reveals that price discovery is more systematic and explainable in the corn market and most related to trading volume. The share of price discovery is also related to time to expiration, especially in the corn market. Other factors such as backwardation, USDA reports, and market crashes influence price discovery, but their effects are much less important. The evidence also shows that price discovery in corn and live cattle futures respond differently to these factors which can be attributed to differences in storability as well as other market characteristics.

Overall, despite dramatic differences in market environment and in the speed and methods of exchange, our findings are highly consistent with the work of earlier, well-respected studies of agricultural futures markets (e.g., Working 1948, 1949; Tomek 1997; Leuthold et al. 1989). Future research could expand our analysis to other markets, examine the relationships at more disaggregate temporal units and intervals within the day, and focus more specifically on how the intraday price discovery process changes on USDA announcement days. Such efforts complement the work by Adjemian (2012), Lehecka, Wang, and Garcia (2014), and Joseph and Garcia (2017).

**Footnotes**

[1] Following Putniņš (2013), we attribute *CS* to the Harris–McInish–Wood component share. However, this measure was introduced by Gonzalo and Granger (1995) and others earlier.

[2] Other reasons include short-run market inefficiency (Schroeder and Goodwin, 1991), pricing mechanisms of the nearby contract being altered by delivery conditions (Garcia, Irwin, and Smith 2015; Leuthold 1972), and prices for nearby and deferred futures contracts reflecting differing information.

[3] Stationarity tests were also conducted. While stationarity and Johansen rank tests often give similar results, they can differ. Recent research by Reed and Smith (2017) demonstrates unit root tests are unreliable in the presence of cointegration. As a result, we follow Fricke and Menkhoff (2011) who categorize their data based on results from Johansen rank test.

[4] To save space, details are presented in an online appendix.

[5] Examination over time of the individual price discovery shares just prior to expiration, revealed that the average values reported in figure 6 are not driven by outliers or an abundance of non-cointegration days. Rather they appear to be reflecting long-observed intertemporal volatility that occurs at expiration.

[6] The overall effect of days to expiration based on the coefficient estimates is $-0.008 Expiration_d + 0.00033 Expiration^2{}_d$ which is greater than 0 when $Expiration_d$ is greater than 24.

Fricke, C. and Menkhoff, L. 2011. "Does the "Bund" Dominate Price Discovery in Euro Bond Futures? Examining Information Shares." *Journal of Banking & Finance* 35(5): 1057-1072.

Garbade, K. and Silber, W. 1983. "Price Movements and Price Discovery in Futures and Cash Markets." *The Review of Economics and Statistics* 65(2): 289-297.

Garcia, P., Irwin, S.H. and Smith, A. 2015. "Futures Market Failure?" *American Journal of Agricultural Economics* 97(1): 40-64.

Garcia, P. and R.M. Leuthold. 1992. "The Effect of Market Information on Corn and Soybean Markets." Proceedings of NCR-134 Conference on Applied Commodity Price Analysis, Forecasting, and Market Risk Management, Chicago, IL.

Gee, K. 2016. "Welcome to the Meat Casino! The Cattle Futures Market Descends into Chaos." *Wall Street Journal*, August 17, pp. 1.

Geoffrey Booth, G. and Ciner, C. 2001. "Linkages among Agricultural Commodity Futures Prices: Evidence from Tokyo." *Applied Economics Letters 8*(5):311-313.

Gonzalo, J. and Granger, C. 1995. "Estimation of Common Long-Memory Components in Cointegrated Systems." *Journal of Business & Economic Statistics* 13(1): 27-35.

Gousgounis, E., and E. Onur. 2016. "The Effect of Pit Closure on Futures Trading." Paper presented at the NCCC-134 Conference on Applied Commodity Price Analysis, Forecasting, and Market Risk Management, St. Louis, MO. April 18-19.

Gray, R.W. and Rutledge, D.J. 1971. "The Economics of Commodity Futures Markets: A survey." *Review of Marketing and Agricultural Economics* 39(4): 57-108.
32

Koontz, S., P. Garcia and M. Hudson (1990). Dominant-Satellite Relationships in Live Cattle Markets. *Journal of Futures Markets* 10: 123-136.

Leath, M.N. and Garcia, P. 1983. "The Efficiency of the Corn Futures Market in Establishing Forward Prices." *North Central Journal of Agricultural Economics* 5(2): 91-101.

Lehecka, G. V. 2014. "The Value of USDA Crop Progress and Condition Information: Reactions of Corn and Soybean Futures Markets." *Journal of Agricultural and Resource Economics* 39(1): 88-105.

Lehecka, G.V., Wang, X. and Garcia, P. 2014. "Gone In Ten Minutes: Intraday Evidence of Announcement Effects in the Electronic Corn Futures Market." *Applied Economic Perspectives and Policy* 36(3): 504-526.

Leuthold, R.M. 1972. "Random Walk and Price Trends: The Live Cattle Futures Market." *The Journal of Finance* 27(4): 879-889.

Leuthold, R.M., Junkus, J.C. and Cordier, J.E. 1989. *The theory and practice of futures markets*. Champaign: Stipes Publishing.

Liu, S.M., Brorsen, B.W., Oellermann, C.M. and Farris, A.L. 1994. "Forecasting the Nearby Basis of Live Cattle." *Journal of Futures Markets* 14(3): 259-273.

Mattos, F.L. and Silveira, R.L. 2016. "Futures Price Response to Crop Reports in Grain Markets." *Journal of Futures Markets* 36(10): 923-942.

Mizrach, B. and Neely, C.J. 2008. "Information Shares in the US Treasury Market." *Journal of Banking & Finance* 32(7): 1221-1233.
34

Tomek, W.G. 1997. "Commodity Futures Prices as Forecasts." *Review of Agricultural Economics* 19(1): 23-44.

Weaver, R. D., and Banerjee, A. 1982. Cash Price Variation in the Live Beef Cattle Market: The Causal Role of Futures Trade. *Journal of Futures Markets*, *2*(4):367-389.

Working, H. 1948. "Theory of the Inverse Carrying Charge in Futures Markets." *Journal of Farm Economics*, 30(1):1-28.

Working, H. 1949. "The Theory of Price of Storage." *The American Economic Review* 39(6): 1254-1262.

Yan, B. and Zivot, E. 2010. "A Structural Analysis of Price Discovery Measures." *Journal of Financial Markets*, 13(1):1-19.

Yang, J., Bessler, D., and Leatham, D. 2001. "Asset Storability and Price Discovery in Commodity Futures Markets: A New Look." *The Journal of Futures Markets*, 21(2): 279-300.
36

**Table 1. Distribution of Days Based on Johansen Rank Test Results, 2008-2015**

|  | Stationarity | Cointegration | Non-cointegration | Total |
|---|---|---|---|---|
| Corn |  |  |  |  |
| Nearby and Deferred 1 | 27.86% | 68.93% | 3.21% | 1960 |
| Nearby and Deferred 2 | 23.93% | 71.17% | 4.90% | 1960 |
| Nearby and Deferred 3 | 21.38% | 73.83% | 4.80% | 1960 |
| Nearby and Deferred 4 | 22.13% | 72.39% | 5.48% | 1952 |
|  |  |  |  |  |
| Live Cattle |  |  |  |  |
| Nearby and Deferred 1 | 20.87% | 71.85% | 7.28% | 1950 |
| Nearby and Deferred 2 | 19.32% | 70.79% | 9.89% | 1931 |
| Nearby and Deferred 3 | 18.63% | 67.64% | 13.74% | 1820 |

Note: Results are based on Johansen rank hypothesis tests between intraday nearby and deferred futures prices at the 5% significance level using trace statistics. Percentages are given as percentage of total number of days.

**Table 2. Average Price Discovery and Volume Shares for the Nearby Contract, 2008 -2015**

| Contract Pair | ILS | GS | Price Discovery Share | Volume Share |
|---|---|---|---|---|
| Corn | | | | |
| Nearby vs Deferred 1 | 0.509 | 0.554 | 0.522 | 0.532 |
| Nearby vs Deferred 2 | 0.643 | 0.705 | 0.658 | 0.700 |
| Nearby vs Deferred 3 | 0.728 | 0.750 | 0.733 | 0.797 |
| Nearby vs Deferred 4 | 0.776 | 0.778 | 0.776 | 0.855 |
| | | | | |
| Live cattle | | | | |
| Nearby vs Deferred 1 | 0.423 | 0.402 | 0.418 | 0.332 |
| Nearby vs Deferred 2 | 0.568 | 0.543 | 0.562 | 0.504 |
| Nearby vs Deferred 3 | 0.641 | 0.610 | 0.635 | 0.621 |

Note: Days with cointegrated $I(1)$ intraday nearby and deferred prices are used for the *ILS*. Days with stationary intraday nearby and deferred prices are used for the *GS*. Price discovery share represents the combination of *ILS* and *GS* estimates.

**Table 3. Regression Results for Corn Futures Contracts**

|  | Nearby and Deferred 1 | Nearby and Deferred 2 | Nearby and Deferred 3 | Nearby and Deferred 4 |
|---|---|---|---|---|
| *Intercept* | 0.030** | 0.056** | 0.090*** | 0.101** |
|  | (0.014) | (0.023) | (0.030) | (0.042) |
| *Volumeshare* | 0.733*** | 0.615*** | 0.644*** | 0.681*** |
|  | (0.024) | (0.040) | (0.043) | (0.056) |
| *Expiration* | 0.006*** | 0.010*** | 0.008*** | 0.006*** |
|  | (0.001) | (0.002) | (0.002) | (0.002) |
| *Expiration$^2$* | 0.000*** | 0.000*** | 0.000*** | 0.000*** |
|  | (0.000) | (0.000) | (0.000) | (0.000) |
| *Backwardation* | 0.069*** | 0.006 | -0.005 | -0.004 |
|  | (0.013) | (0.017) | (0.013) | (0.012) |
| *WASDE & CP* | -0.007 | -0.026 | -0.065*** | -0.039 |
|  | (0.013) | (0.017) | (0.021) | (0.024) |
| *Grainstocks* | 0.000 | 0.006 | -0.024 | -0.102** |
|  | (0.035) | (0.041) | (0.038) | (0.049) |
| *Crash* | -0.015 | 0.048 | 0.065*** | 0.045* |
|  | (0.017) | (0.031) | (0.019) | (0.024) |
| *Stationarity* | 0.014* | 0.037*** | 0.011 | -0.006 |
|  | (0.008) | (0.010) | (0.011) | (0.011) |
| Adjusted $R^2$ | 0.781 | 0.693 | 0.638 | 0.517 |
| Observations | 1897 | 1864 | 1866 | 1845 |

Note: This table reports regression results in which the estimated price discovery share for corn is the dependent variable for all equations. *Volumeshare* is the nearby contract's volume share. *Expiration* is the number of days to the nearby contract's expiration. *Backwardation* is a dummy variable equals to one if deferred futures settlement price was below nearby futures settlement price and zero otherwise. *WASDE CP* is a dummy variable equals to one on USDA WASDE and Corp Production report days, and zero otherwise. *Grainstocks* is a dummy variable equals to one on USDA Grain Stocks report days and zero otherwise. *Crash* is a dummy variable equals to one in the period of July 03, 2008 to December 08, 2008 and zero otherwise. *Stationarity* is a dummy variable equals to one if intraday nearby and deferred prices were both stationary, and zero otherwise. Newey-West standard errors are reported in parenthesis. Numbers are rounded to the third decimal place. Asterisks ***, **, and * indicate significance at the 1%, 5%, and 10% levels, respectively.

**Table 4. Regression Results for Live Cattle Futures Contracts**

|  | Nearby and Deferred 1 | Nearby and Deferred 2 | Nearby and Deferred 3 |
|---|---|---|---|
| *Intercept* | 0.295*** | 0.243*** | 0.291*** |
|  | (0.026) | (0.026) | (0.033) |
| *Volumeshare* | 0.300*** | 0.388*** | 0.380*** |
|  | (0.080) | (0.078) | (0.077) |
| *Expiration* | -0.008*** | 0.005 | 0.006 |
|  | (0.003) | (0.003) | (0.003) |
| *Expiration$^2$* | 0.000*** | 0.000 | 0.000 |
|  | (0.000) | (0.000) | (0.000) |
| *Backwardation* | -0.015 | 0.010 | 0.011 |
|  | (0.013) | (0.016) | (0.018) |
| *CF* | 0.035 | -0.007 | -0.043 |
|  | (0.028) | (0.030) | (0.033) |
| *Crash* | -0.044* | -0.061* | -0.069*** |
|  | (0.025) | (0.032) | (0.026) |
| *Stationarity* | -0.024** | -0.021 | -0.032** |
|  | (0.011) | (0.013) | (0.014) |
| Adjusted $R^2$ | 0.294 | 0.278 | 0.224 |
| Observations | 1808 | 1740 | 1570 |

Note: This table reports regression results in which the estimated price discovery share for live cattle is the dependent variable for all equations. *Volumeshare* is the nearby contract's volume share. *Expiration* is the number of days to the nearby contract's expiration. *Backwardation* is a dummy variable equals to one if deferred futures settlement price was below nearby futures settlement price and zero otherwise. *CF* is a dummy variable equals to one on the trading day following the release of Cattle on Feed report. *Crash* is a dummy variable equals to one in the year of 2015. *Stationarity* is a dummy variable equals to one if intraday nearby and deferred prices were both stationary, and zero otherwise. Newey-West standard errors are reported in parenthesis. Numbers are rounded to the third decimal place. Asterisks ***, **, and * indicate significance at the 1%, 5%, and 10% levels, respectively.

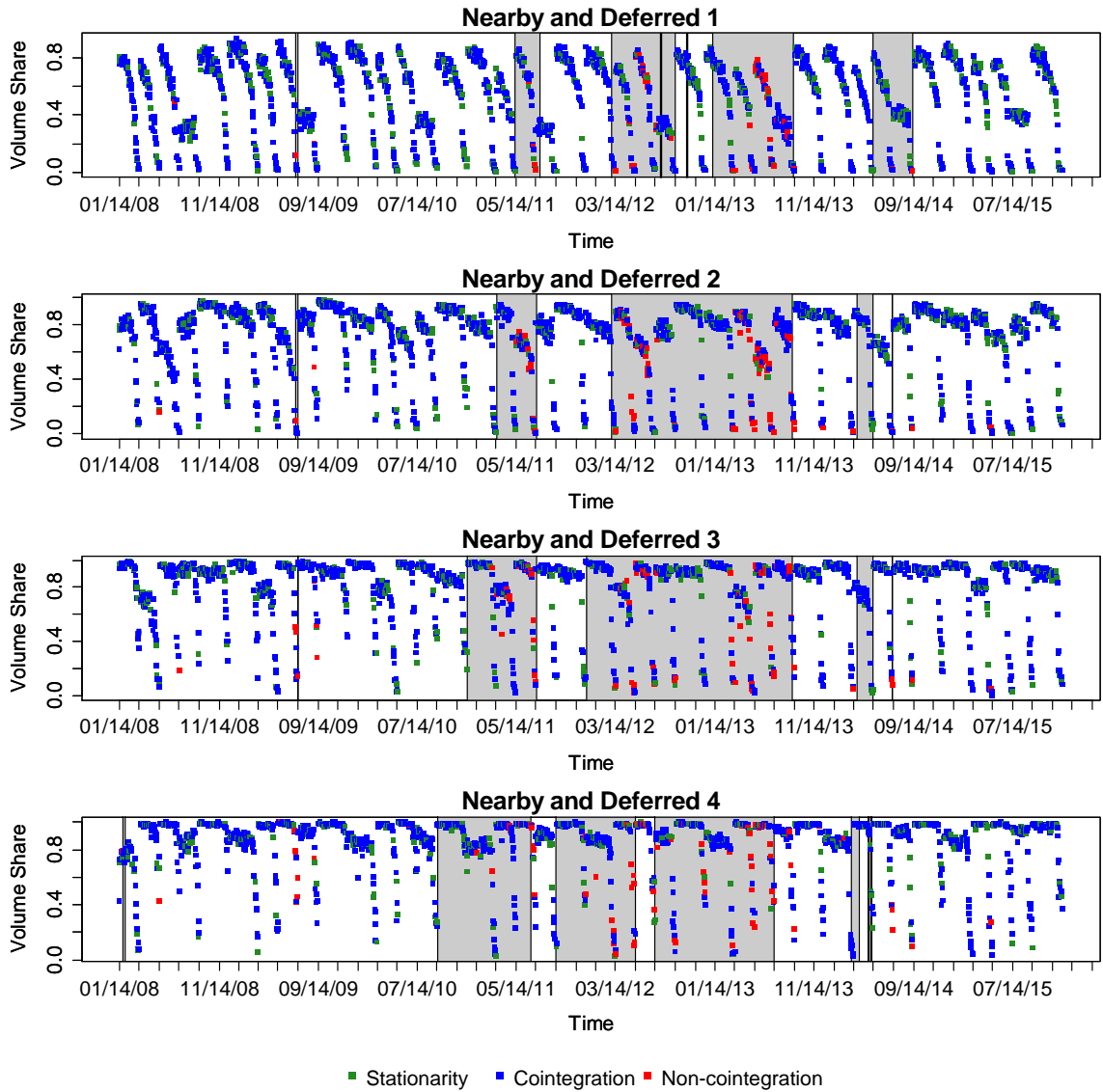

**Figure 1. Volume shares and Johansen rank test results for corn futures, 2008-2015**

Note: Shaded areas represent backwardation periods. Corn futures contracts expire on the business day prior to the 15th calendar day of the maturity month.

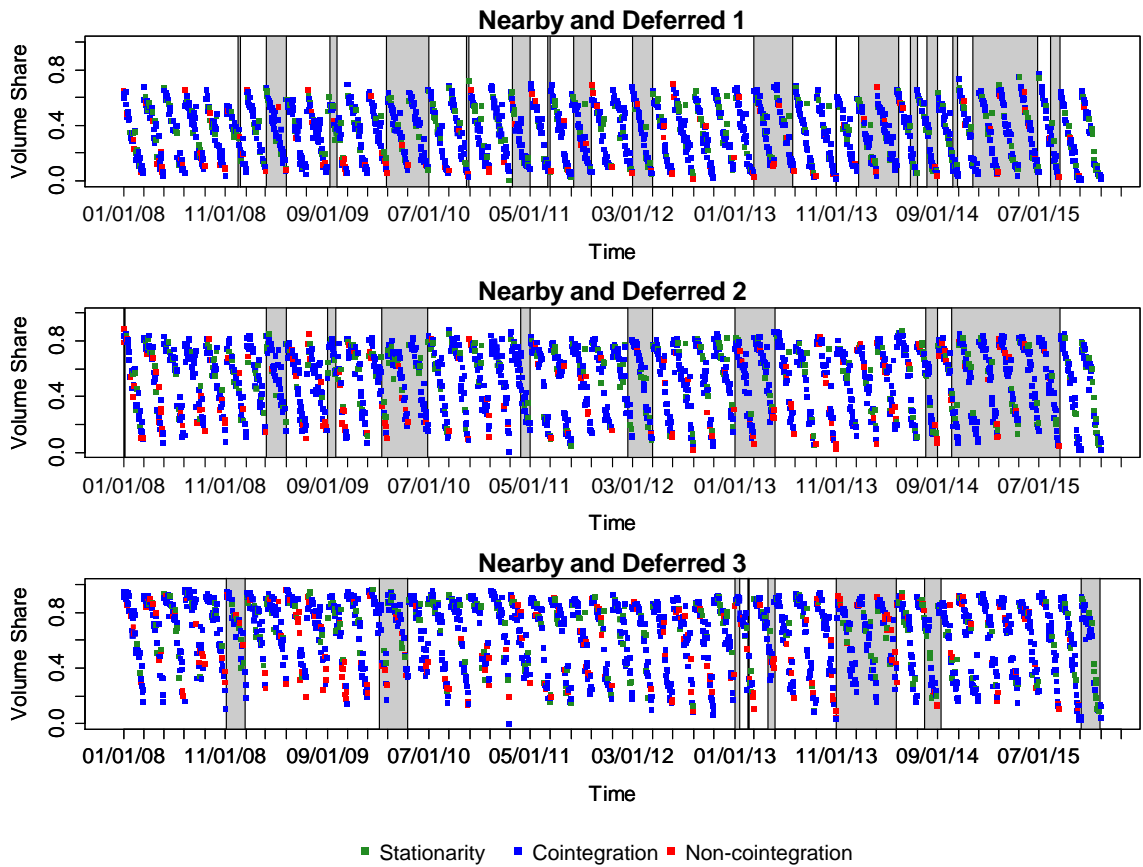

**Figure 2. Volume shares and Johansen rank test results for live cattle futures, 2008-2015**

Note: Shaded areas represent backwardation periods. Live cattle futures contracts expire on the last business day of the maturity month.

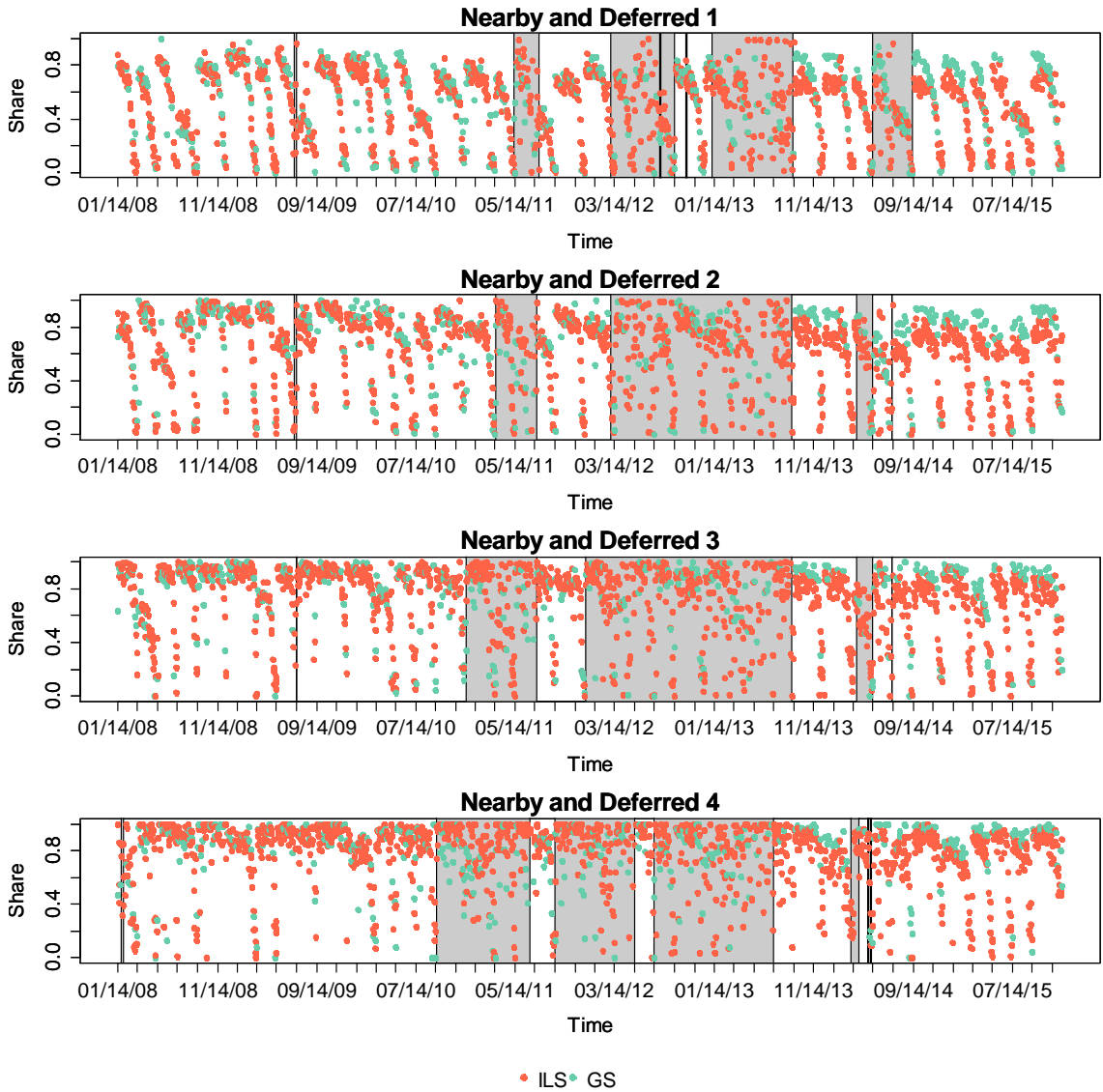

**Figure 3. Price discovery shares for the nearby contract compared to deferred 1, 2, 3, and 4 contracts in the corn futures market, 2008-2015**

Note: Shaded areas represent backwardation periods. Corn futures contracts expire on the business day prior to the 15th calendar day of the maturity month.

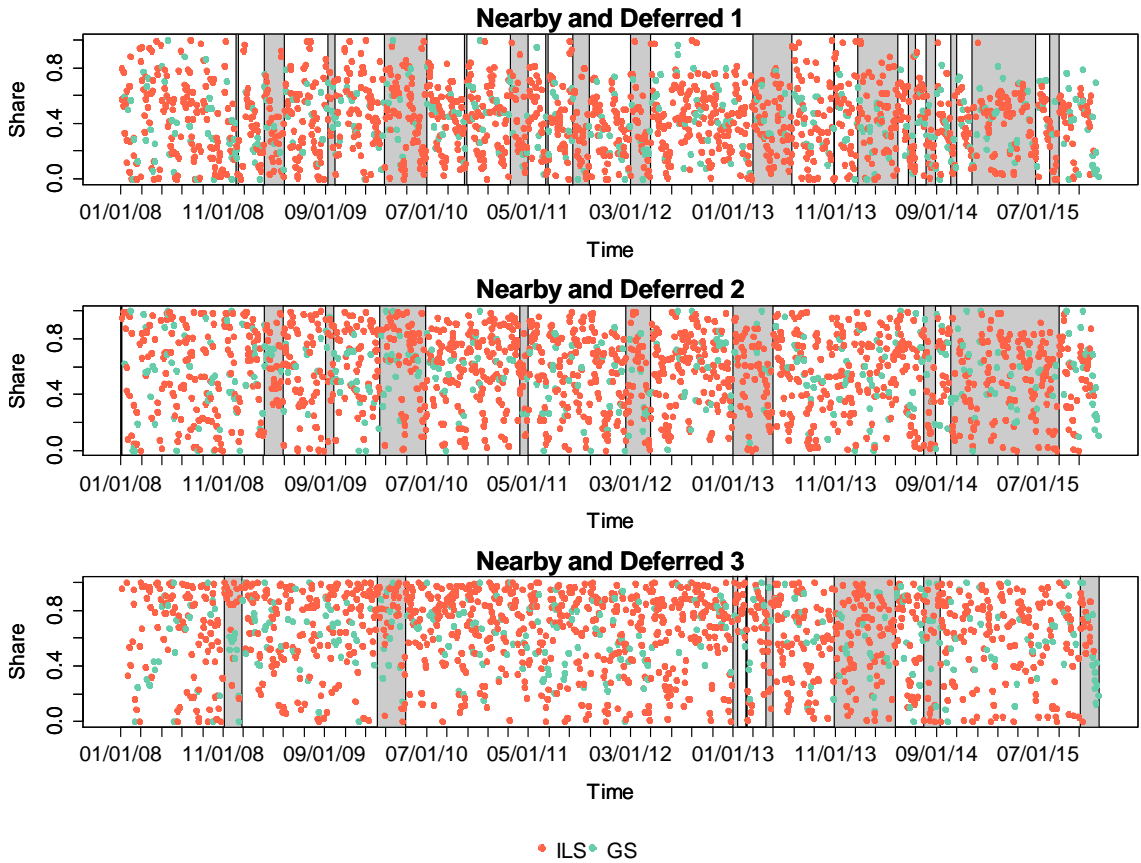

**Figure 4. Price discovery shares for the nearby contract compared to deferred 1, 2, and 3 contracts in the live cattle futures market, 2008-2015**

Note: Shaded areas represent backwardation periods. Live cattle futures contracts expire on the last business day of the maturity month.

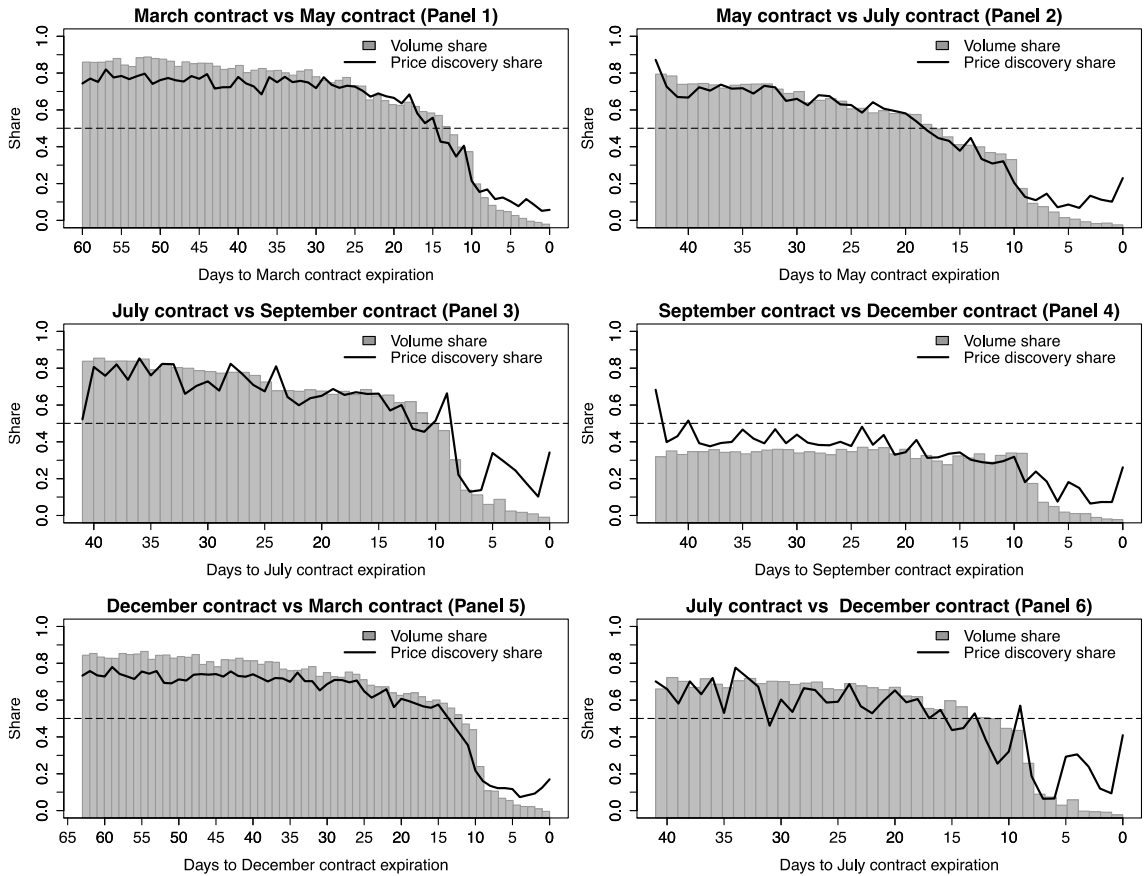

**Figure 5. Price discovery and volume shares in the nearby period for each contract month in the corn futures market**

Note: Panels show the average over years of volume share of nearby relative to the first deferred and *ILS* between nearby and first deferred contract. The information is organized along the x-axis by days to maturity of the nearby.

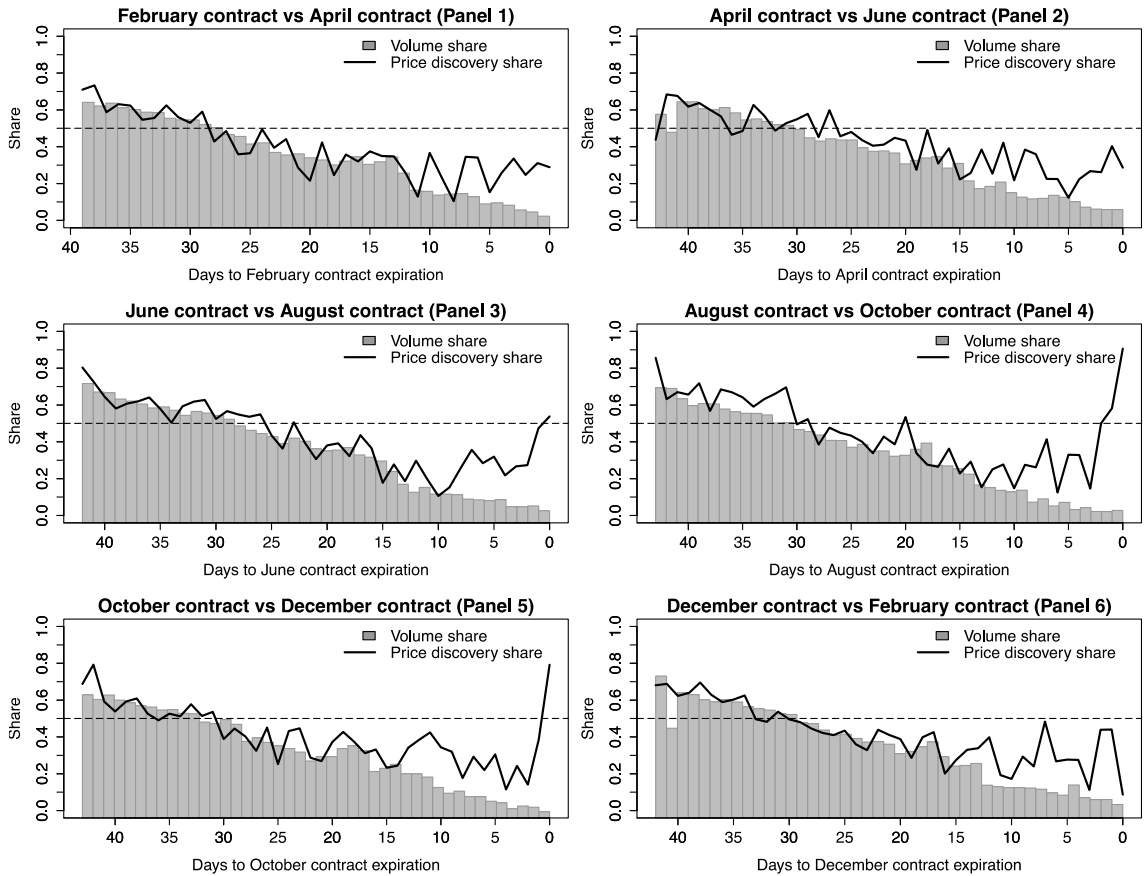

**Figure 6. Price discovery and volume shares in the nearby period for each contract month in the live cattle futures market**

Note: Panels show the average over years of volume share of nearby relative to the first deferred and *ILS* between nearby and first deferred contract. The information is organized along the x-axis by days to maturity of the nearby.

# APPENDIX A

**This appendix provides results for footnote 2 in the article.**

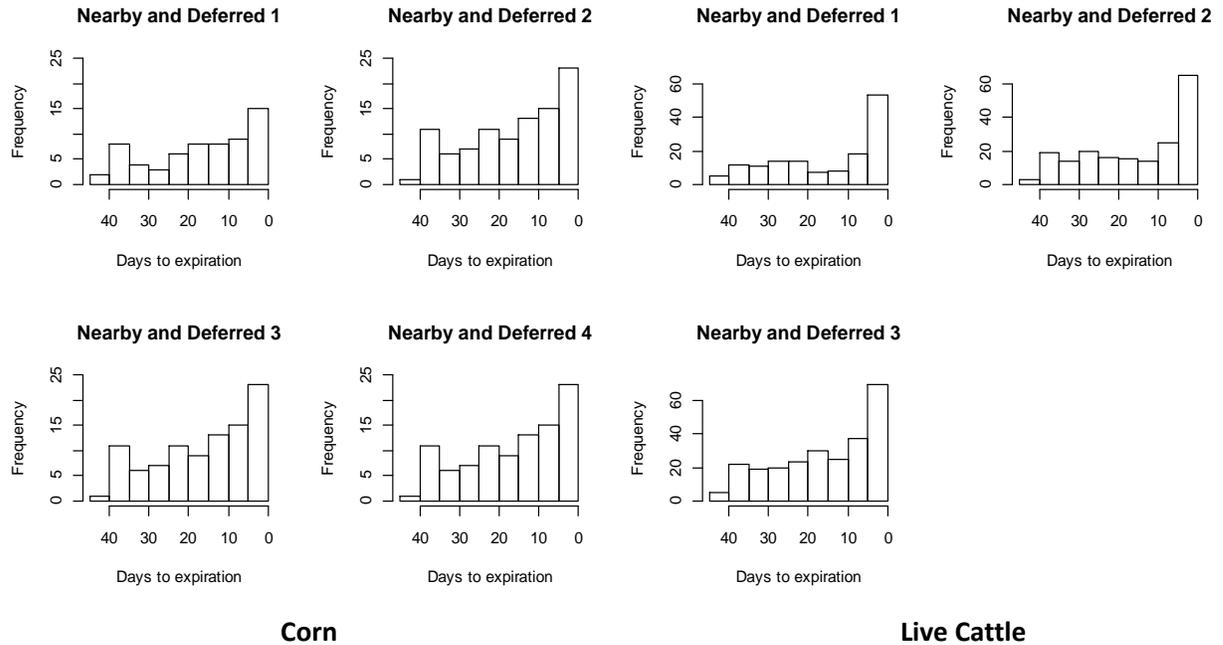

**Corn**            **Live Cattle**

**Figure 1. Histograms of days when intraday nearby and deferred futures prices were not cointegrated**

Note: Histograms for corn and live cattle are in the left and right panels, respectively.

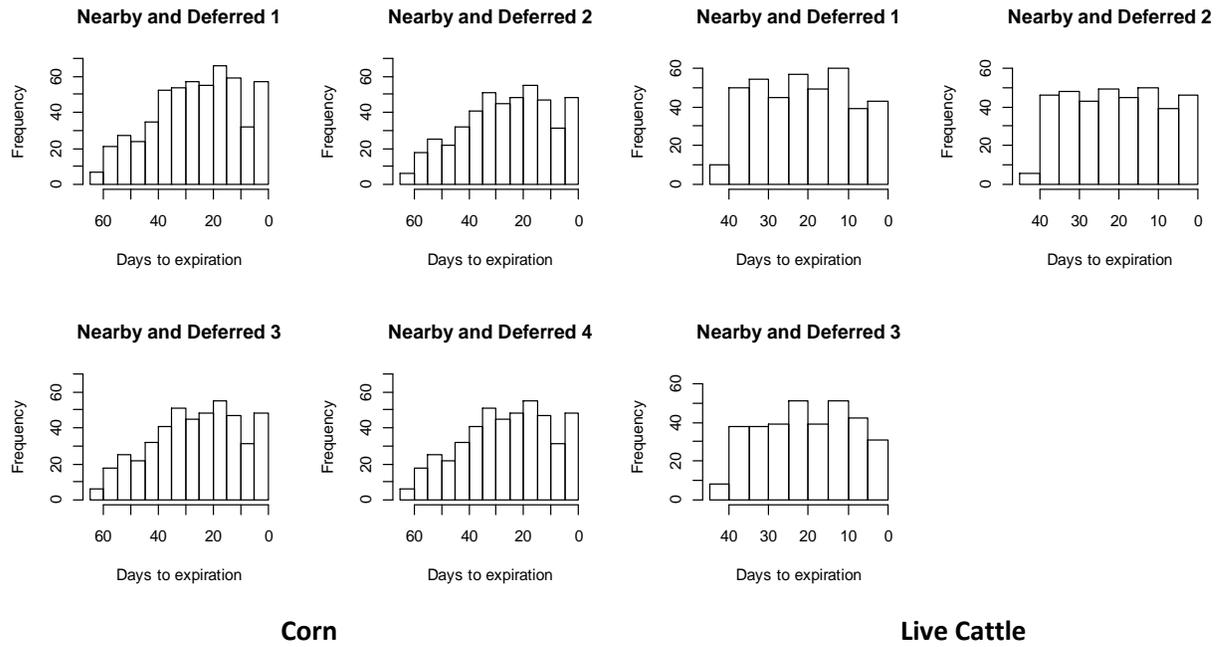

**Figure 2. Histograms of days when intraday nearby and deferred futures prices were both stationary**

Note: Histograms for corn and live cattle are in the left and right panels, respectively.

# APPENDIX B

**This appendix provides daily average price discovery and volume shares for individual contracts in nearby periods.**

**Table 1. Daily Average Price Discovery and Volume Shares for Individual Contracts in Nearby Periods, 2008 –2015**

| Contract Month | Price Discovery Share | Volume Share |
|---|---|---|
| Corn | | |
| March | 0.592 | 0.623 |
| May | 0.478 | 0.477 |
| July | 0.582 | 0.568 |
| September | 0.331 | 0.293 |
| December | 0.583 | 0.631 |
| | | |
| Live cattle | | |
| February | 0.399 | 0.369 |
| April | 0.425 | 0.363 |
| June | 0.426 | 0.368 |
| August | 0.438 | 0.355 |
| October | 0.407 | 0.350 |
| December | 0.410 | 0.359 |